# Entropic Effects of Thermal Rippling on van der Waals Interactions between Monolayer Graphene and a Rigid Substrate

Peng Wang[1], Wei Gao[2], and Rui Huang[1*]

[1]*Department of Aerospace Engineering and Engineering Mechanics, University of Texas, Austin, TX 78712, USA*

[2]*Department of Mechanical Engineering, Northwestern University, Evanston, IL 60208, USA*

**Abstract**

Graphene monolayer, with extremely low flexural stiffness, displays spontaneous rippling due to thermal fluctuations at a finite temperature. When a graphene membrane is placed on a solid substrate, the adhesive interactions between graphene and the substrate could considerably suppress thermal rippling. On the other hand, the statistical nature of thermal rippling adds an entropic contribution to the graphene-substrate interactions. In this paper we present a statistical mechanics analysis on thermal rippling of monolayer graphene supported on a rigid substrate, assuming a generic form of van der Waals interactions between graphene and substrate at $T = 0$ K. The rippling amplitude, the equilibrium average separation, and the average interaction energy are predicted simultaneously and compared with molecular dynamics (MD) simulations. While the amplitude of thermal rippling is reduced by adhesive interactions, the entropic contribution leads to an effective repulsion. As a result, the equilibrium average separation increases and the effective adhesion energy decreases with increasing temperature. Moreover, the effect of a biaxial pre-strain in graphene is considered, and a buckling instability is predicted at a critical compressive strain that depends on both the temperature and the adhesive interactions. Limited by the harmonic approximations, the theoretical predictions agree with MD simulations only for relatively small rippling amplitudes but can be extended to account for the anharmonic effects.

[*] Corresponding author. Email: ruihuang@mail.utexas.edu.

# I. INTRODUCTION

Graphene and other two-dimensional (2D) materials have drawn extensive interests for research due to their remarkable structures and properties. One of the common features among these 2D materials is their monatomic thickness. As a result, they are highly flexible with extremely low flexural rigidity, compared to conventional membranes and thin film materials. At a finite temperature (T > 0 K), thermal fluctuations of such ultrathin membranes are expected [1, 2], similar to the ubiquitous fluctuations of biomembranes [3-5]. Indeed, experimental observations have found that suspended graphene membranes often display spontaneous ripples [1, 6, 7], likely a result of thermal fluctuations [2]. Such thermal rippling has been found to be responsible for the temperature dependent mechanical properties of graphene including elastic modulus and apparently negative coefficient of thermal expansion (CTE) at the room temperature [8-10]. In most applications, graphene membranes are supported on solid substrates, such as silicon (with an oxide surface), copper, and polymers. In addition to the intrinsic thermal rippling, the morphology of a substrate-supported graphene membrane depends on the surface roughness of the substrate as well as the interactions between graphene and the substrate. Ripples, wrinkles and folds are commonly observed in supported graphene as well as other 2D materials [11-16]. Many physical properties of graphene depend on the morphology that may be altered by the interactions with a substrate. In this paper, we present a statistical mechanics analysis on thermal rippling of monolayer graphene supported on a rigid substrate and corresponding molecular dynamics simulations for comparison. Two main questions are to be answered: First, how would the rippling morphology depend on the adhesive interactions? Second, how would the statistical thermal rippling influence the graphene-substrate interactions at a finite temperature?

The mechanisms of adhesive interactions between graphene and typical substrates such as silicon oxide ($SiO_2$) and metals have been studied recently. Both experiments [17-21] and first-principle calculations [22, 23] have suggested that van der Waals interactions are the primary mechanisms in most cases, although other mechanisms may also exist in some cases [24-26]. In the present study, we assume a generic form of van der Waals interactions between graphene and the substrate at $T = 0$ K, which was derived from the Lennard-Jones (LJ) potential for pairwise particle-particle interactions [27]. Such an adhesive interaction is expected to suppress the rippling amplitude of a supported graphene membrane. However, a quantitative correlation between adhesion and rippling morphology of graphene has yet to be established. Moreover, even with temperature-independent parameters for the van der Waals interactions, the statistical nature of thermal rippling renders an entropic effect on the graphene-substrate interactions that would depend on temperature. As a result, the effective properties of the graphene-substrate interface become temperature dependent in general. Furthermore, additional effects on the morphology and adhesion of graphene may come from the fact that the graphene membrane is often subjected to an in-plane pre-strain, either unintentionally due to the growth/transfer processes or intentionally for the purpose of strain engineering [28].

The remainder of this paper is organized as follows. Section II presents a statistical mechanics analysis based on a continuum membrane model of pre-strained graphene and the generic form of van der Waals interactions. Section III describes the MD simulations. The results are compared and discussed in Section IV, followed by a summary in Section V.

## II. A CONTINUUM STATISTICAL MECHANICS ANALYSIS

The graphene monolayer is modeled as a two-dimensional (2D) continuum membrane, which interacts with the substrate via an interfacial force field of van der Waals type. The presence of an interfacial force field influences thermal rippling of graphene, which in turn introduces an entropic effect on the graphene-substrate interactions at a finite temperature. The substrate is assumed to be rigid with a perfectly flat surface, whereas the effect of surface roughness is left for future studies.

For a graphene monolayer on a perfectly flat substrate with no thermal rippling, a generic form of the van der Waals interaction energy function can be written as [27]:

$$V(z) = \Gamma_0 \left[ \frac{1}{2} \left( \frac{h_0}{z} \right)^9 - \frac{3}{2} \left( \frac{h_0}{z} \right)^3 \right], \quad (1)$$

where $z$ is the separation distance between graphene and the substrate surface, $\Gamma_0$ is the adhesion energy (per unit area), and $h_0$ is the equilibrium separation. The two parameters ($\Gamma_0$ and $h_0$) are assumed to be independent of temperature in the present study, although they could be temperature dependent in principle (e.g., due to statistical effects of electromagnetic modes and thermal radiation [29-32]).

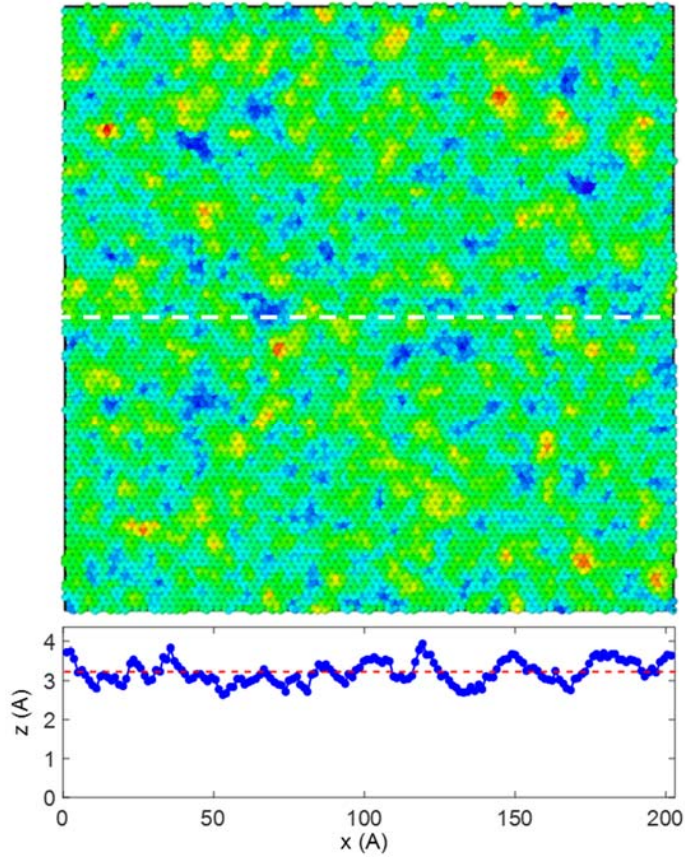

Figure 1. Thermal rippling of graphene on a rigid substrate by MD simulation ($\Gamma_0 = 0.242$ J/m$^2$, $h_0 = 0.316$ nm, $\varepsilon_0 = 0$, and $T = 1000$ K): a top-view snapshot with color contour for the height and a deflection profile along a line.

At a finite temperature ($T > 0$ K), the graphene membrane fluctuates out of the plane (see Fig. 1). At a particular instance, the rippling profile of the graphene can be written as

$$z(x, y; T) = [\bar{z}(T) + w(x, y; T)] h_0 \qquad (2)$$

where $\bar{z}$ and $w$ are the normalized average separation and out-of-plane deflection, respectively. Correspondingly, the total interaction energy between graphene and the substrate over an area $\Omega$ is approximately

$$U_I = \iint_\Omega V(z) dx dy \approx \iint_\Omega \left[ V(\bar{z} h_0) + V'(\bar{z} h_0) h_0 w + \frac{1}{2} V''(\bar{z} h_0) h_0^2 w^2 \right] dx dy, \qquad (3)$$

where $V'(z)$ and $V''(z)$ are the first and second derivatives of the interaction energy function in Eq. (1), and the higher order terms are neglected. Note that this approximation is valid only when $|w| \ll 1$ (i.e., the out-of-plane deflection is small compared to the equilibrium separation).

Following a previous work for freestanding graphene [10], we consider a graphene membrane subjected to a biaxial pre-strain $\varepsilon_0$, relative to the ground state at 0 K. With the rippling profile in Eq. (2), the elastic strain energy of graphene consists of two parts, the bending energy ($U_b$) and in-plane strain energy ($U_s$):

$$U_b \approx \frac{Dh_0^2}{2} \iint_\Omega \left(\frac{\partial^2 w}{\partial x^2} + \frac{\partial^2 w}{\partial y^2}\right)^2 dxdy, \qquad (4)$$

$$U_s \approx \iint_\Omega \left\{ E^* \varepsilon_0^2 + \frac{E^* \varepsilon_0 h_0^2}{2}\left[\left(\frac{\partial w}{\partial x}\right)^2 + \left(\frac{\partial w}{\partial y}\right)^2\right]\right\} dxdy, \qquad (5)$$

where $D$ is the bending modulus of graphene, $E^* = E/(1-\nu)$ is the in-plane biaxial modulus, and $E$ and $\nu$ are the 2D Young's modulus (unit: N/m) and Poisson's ratio of graphene. Note that the bending energy due to Gaussian curvature has been ignored in Eq. (4) and only the quadratic terms of the deflection are retained in Eq. (5) for a harmonic approximation, as discussed in [10].

Assuming periodic boundary conditions in the $x$-$y$ plane, the deflection $w(x, y)$ can be written in form of the Fourier series:

$$w(\mathbf{r}) = \sum_k \hat{w}(\mathbf{q}_k) e^{i\mathbf{q}_k \cdot \mathbf{r}}, \qquad (6)$$

and the corresponding Fourier coefficients are

$$\hat{w}(\mathbf{q}_k) = \frac{1}{L_0^2} \iint_\Omega w(\mathbf{r}) e^{-i\mathbf{q}_k \cdot \mathbf{r}} dxdy, \qquad (7)$$

where $\mathbf{r}$ is the 2D position vector, $\mathbf{q}_k$ denotes the $k$-th wave vector in the 2D space, and $L_0^2$ is the area of the domain $\Omega$. For each configuration, the mean-square amplitude of the out-of-plane fluctuation is then

$$\delta^2 = \frac{h_0^2}{L_0^2} \iint_\Omega w^2(\mathbf{r}) dxdy = h_0^2 \sum_k \left(\hat{w}_{\text{Re}}^2(\mathbf{q}_k) + \hat{w}_{\text{Im}}^2(\mathbf{q}_k)\right), \qquad (8)$$

where $\hat{w}_{\text{Re}}(\mathbf{q}_k)$ and $\hat{w}_{\text{Im}}(\mathbf{q}_k)$ are the real and imaginary parts of $\hat{w}(\mathbf{q}_k)$, respectively.

Considering the statistical nature of thermal rippling, the Fourier coefficients $\hat{w}_{\text{Re}}(\mathbf{q}_k)$ and $\hat{w}_{\text{Im}}(\mathbf{q}_k)$ are taken as continuous random variables. Each set of $\hat{w}_{\text{Re}}(\mathbf{q}_k)$ and $\hat{w}_{\text{Im}}(\mathbf{q}_k)$ constitutes a possible configuration of the membrane. All possible configurations of the membrane construct a statistical ensemble. Based on classical statistical mechanics [33, 34], the probability density function (PDF) for each configuration is given by Boltzmann distribution at thermal equilibrium:

$$\rho = \frac{1}{Z}\exp\left(-\frac{U}{k_B T}\right), \tag{9}$$

where $U$ is the total potential energy of the configuration, $Z$ is the configurational partition function, and $k_B$ is Boltzmann constant. Substituting Eq. (6) into Eqs. (3-5), the total potential energy for each configuration of the supported graphene membrane is obtained in terms of the Fourier coefficients as

$$\begin{aligned} U &= U_I + U_b + U_s \\ &\approx L_0^2\left(V(\bar{z}h_0) + E^*\varepsilon_0^2\right) + L_0^2 h_0^2 \sum_{k(\mathbf{q}_k \cdot \mathbf{e}_y \geq 0)} \left(Dq_k^4 + E^*\varepsilon_0 q_k^2 + V''(\bar{z}h_0)\right)\left(\hat{w}_{\text{Re}}^2(\mathbf{q}_k) + \hat{w}_{\text{Im}}^2(\mathbf{q}_k)\right), \end{aligned} \tag{10}$$

where $q_k = |\mathbf{q}_k|$ is the amplitude of the wave vector. It is important to note that the coefficients $\hat{w}(\mathbf{q}_k)$ and $\hat{w}(-\mathbf{q}_k)$ are not independent since the deflection in Eq. (6) must be real valued. Consequently, only those Fourier coefficients associated with the upper half-plane of the wave vectors (i.e., $\mathbf{q}_k \cdot \mathbf{e}_y \geq 0$, including only half of the *x*-axis) are taken as the independent random variables in Eq. (10).

By the equipartition theorem [34], the mean energy associated with each independent harmonic term in Eq. (10) equals $k_B T/2$, and thus we obtain that

$$\left\langle \hat{w}_{\text{Re}}^2(\mathbf{q}_k) \right\rangle = \left\langle \hat{w}_{\text{Im}}^2(\mathbf{q}_k) \right\rangle = \frac{k_B T}{2L_0^2 h_0^2\left(Dq_k^4 + E^*\varepsilon_0 q_k^2 + V''(\bar{z}h_0)\right)}. \tag{11}$$

where $\langle \cdot \rangle$ denotes the ensemble average of the enclosed quantity. The ensemble average of the mean-square amplitude in Eq. (8) is then

$$\left\langle \delta^2 \right\rangle = \frac{k_B T}{L_0^2} \sum_k \frac{1}{Dq_k^4 + E^*\varepsilon_0 q_k^2 + V''(\bar{z}h_0)}. \tag{12}$$

Without the double derivative of the interaction energy function, Eq. (12) recovers the classical results for undulations of fluid membranes by Helfrich and Servuss [35], and the same result was obtained for a freestanding graphene membrane [10]. The additional term due to the interactions between graphene and the substrate depends on the average separation $\bar{z}h_0$, which is unknown *a priori*. As shown later, the average separation at thermal equilibrium can be determined as a function of the temperature by minimizing the Helmholtz free energy of the graphene/substrate system under the isothermal condition. We note that, for the amplitude in Eq. (12) to be positive definite, it requires that $V''(\bar{z}h_0) \geq 0$ for $\varepsilon_0 \geq 0$ or $4DV''(\bar{z}h_0) > (E^*\varepsilon_0)^2$ for $\varepsilon_0 < 0$, which imposes a limitation for the harmonic approximation in the present analysis.

With Boltzmann distribution in Eq. (9), the configurational partition function for the statitsical thermal rippling is obtained as

$$Z \approx \int_{-\infty}^{\infty} \cdots \int_{-\infty}^{\infty} \exp\left(-\frac{U}{k_B T}\right) d\hat{w}_{\text{Re}}(\mathbf{q}_1) d\hat{w}_{\text{Im}}(\mathbf{q}_1) d\hat{w}_{\text{Re}}(\mathbf{q}_2) d\hat{w}_{\text{Im}}(\mathbf{q}_2) \cdots$$
$$= \exp\left(-\frac{L_0^2}{k_B T}\left(V(\bar{z}h_0) + E^*\varepsilon_0^2\right)\right) \prod_{k(\mathbf{q}_k \cdot \mathbf{e}_y \geq 0)} \left[\left(1 + \frac{E^*\varepsilon_0}{Dq_k^2} + \frac{V''(\bar{z}h_0)}{Dq_k^4}\right)^{-1}\left(\frac{\pi k_B T}{DL_0^2 h_0^2 q_k^4}\right)\right]. \quad (13)$$

Here the integration limits have been taken to be -∞ and ∞ for each random variable. However, the random variables should be limited within a small range ($|w| \ll 1$) under the harmonic approximation. Moreover, the rippling membrane should be constrained so that it does not penetrate into the substrate, which may lead to a steric effect [35-37]. Nevertheless, we proceed with Eq. (13) as an approximate partition function and leave the additional effects for future studies.

With the partition function in Eq. (13), the Helmholtz free energy of the graphene/substrate system is obtained as a function of the average separation, pre-strain and temperature:

$$A(\bar{z}, \varepsilon_0, T) = -k_B T \ln Z$$
$$\approx L_0^2\left(V(\bar{z}h_0) + E^*\varepsilon_0^2\right) - k_B T \sum_{k(\mathbf{q}_k \cdot \mathbf{e}_y \geq 0)} \left[\ln\left(\frac{\pi k_B T}{DL_0^2 h_0^2 q_k^4}\right) - \ln\left(1 + \frac{E^*\varepsilon_0}{Dq_k^2} + \frac{V''(\bar{z}h_0)}{Dq_k^4}\right)\right]. \quad (14)$$

At a given temperature, the Helmholtz free energy can be minimized with respect to the average separation and the pre-strain for the thermomechanical equilibrium state. First, taking derivative of the free energy with respect to the average separation, we obtain the average normal traction (force per unit area) between graphene and the substrate as

$$s(\bar{z},\varepsilon_0,T) = \frac{1}{L_0^2 h_0}\left(\frac{\partial A}{\partial \bar{z}}\right)_{T,\varepsilon_0} = V'(\bar{z}h_0) + \frac{k_B T}{L_0^2}V'''(\bar{z}h_0)\sum_{k(\mathbf{q}_k\cdot\mathbf{e}_y\geq 0)}\left[Dq_k^4 + E^*\varepsilon_0 q_k^2 + V''(\bar{z}h_0)\right]^{-1}, \quad (15)$$

where the first term on the right-hand side is the interfacial traction at 0 K (without thermal rippling) and the second term is the entropic contribution due to thermal rippling. Hence, Eq. (15) predicts a temperature-dependent traction-separation relation for the interactions between graphene and the substrate. The equilibrium average separation, $\bar{z}^*(\varepsilon_0,T)$, is then obtained by setting $s=0$, namely

$$V'(\bar{z}^*h_0) + \frac{k_B T}{L_0^2}V'''(\bar{z}^*h_0)\sum_{k(\mathbf{q}_k\cdot\mathbf{e}_y\geq 0)}\left[Dq_k^4 + E^*\varepsilon_0 q_k^2 + V''(\bar{z}^*h_0)\right]^{-1} = 0. \quad (16)$$

In addition, it is required that $\left(\frac{\partial^2 A}{\partial \bar{z}^2}\right)_{T,\varepsilon_0} > 0$ at $\bar{z}=\bar{z}^*$ for the equilibrium separation to be stable.

Interestingly, we note that, if the interaction energy is purely harmonic with $V'''(z)\equiv 0$, the entropic contribution in Eq. (15) vanishes and the equilibrium average separation becomes independent of temperature ($\bar{z}^*\equiv 1$). In general, however, the interaction energy as given in Eq. (1) is anharmonic, which leads to the entropic effect and the temperature dependence for the equilibrium separation. Therefore, despite the harmonic approximation of the interaction energy function in Eq. (3), the anharmonic effect of the interaction is partly taken into account in Eqs. (15) and (16).

Next, taking derivative of the Helmholtz free energy in Eq. (14) with respect to the pre-strain, we obtain the average in-plane stress (equi-biaxial) in the graphene membrane as

$$\sigma(\bar{z},\varepsilon_0,T) = \frac{1}{2L_0^2}\left(\frac{\partial A}{\partial \varepsilon_0}\right)_{T,\bar{z}} = E^*\left(\varepsilon_0 + \frac{k_B T}{2L_0^2}\sum_{k(\mathbf{q}_k\cdot\mathbf{e}_y\geq 0)}\left[Dq_k^2 + E^*\varepsilon_0 + q_k^{-2}V''(\bar{z}h_0)\right]^{-1}\right). \quad (17)$$

Taking $\bar{z}=\bar{z}^*(\varepsilon_0,T)$, the average in-plane stress at the equilibrium average separation is:

$$\sigma^*(\varepsilon_0,T) = \sigma(\bar{z}^*,\varepsilon_0,T) = E^*\varepsilon_0 + \tilde{\sigma}^*(\varepsilon_0,T), \quad (18)$$

where the first term on the right-hand side is the pre-stress without rippling and the second term is the additional tension due to the entropic effect of thermal rippling (rippling stress):

$$\tilde{\sigma}^*(\varepsilon_0, T) = E^* \frac{k_B T}{2L_0^2} \sum_{k(\mathbf{q}_k \cdot \mathbf{e}_y \geq 0)} \left[ Dq_k^2 + E^* \varepsilon_0 + q_k^{-2} V''(\bar{z}^* h_0) \right]^{-1}. \qquad (19)$$

As noted in the previous study [10], the in-plane thermal fluctuations of the graphene lattice lead to a positive thermal expansion if the out-of-plane fluctuations are completely suppressed. Taking the in-plane thermal expansion into account, the effective in-plane stress in graphene at a finite temperature is approximately

$$\sigma^*(\varepsilon_0, T) \approx E^*(\varepsilon_0 - \alpha_{2D} T) + \tilde{\sigma}^*(\varepsilon_0, T), \qquad (20)$$

where $\alpha_{2D}$ is the 2D coefficient of thermal expansion (2D-CTE) resulting from the anharmonic interactions among in-plane phonon modes and was found to be a constant, $\alpha_{2D} \sim 5.51 \times 10^{-6}$ K$^{-1}$, independent of temperature (up to 1000 K) [10]. Setting $\sigma^*(\varepsilon_0, T) = 0$ in Eq. (20) then leads to an equilibrium thermal strain, $\varepsilon_0^*(T)$, which gives the effective thermal expansion of the supported graphene and could be either positive or negative due to the competing effects between in-plane lattice expansion and out-of-plane rippling, as discussed in the previous studies [8-10] for freestanding graphene.

To be specific, the predictions by the statistical mechanics analysis are illustrated and discussed for a square-shaped graphene membrane. First, the normal traction in Eq. (15) is evaluated by summation over discrete Fourier modes, which can be written in a dimensionless form as

$$\frac{s}{s_0} = f'(\bar{z}) + f'''(\bar{z}) \frac{k_B T}{2D} \sum_{i,j=-n}^{n} \left[ 16\pi^4 \left(\frac{h_0}{L_0}\right)^2 (i^2 + j^2)^2 + 4\pi^2 \beta \varepsilon_0 (i^2 + j^2) + \eta f''(\bar{z}) \left(\frac{L_0}{h_0}\right)^2 \right]^{-1}, \qquad (21)$$

where $s_0 = \Gamma_0 / h_0$, $\eta = \Gamma_0 h_0^2 / D$, $\beta = E^* h_0^2 / D$, $f'(z)$, $f''(z)$, and $f'''(z)$ are derivatives of the normalized interaction energy function, $f(z) = \frac{1}{2}(z^{-9} - 3z^{-3})$. The number $n$ depends on two length scales: the domain size $L_0$ and a microscopic cut-off length $b$ (e.g., the minimum wavelength of thermal rippling). For $L_0 >> b$, $n \to \infty$ and the summation in Eq. (21) converges to a constant. If $\varepsilon_0 = 0$, the summation in Eq. (21) can be calculated by an integral approximation as

$$\frac{s}{s_0} \approx f'(\bar{z}) + f'''(\bar{z}) \frac{\pi k_B T}{D} \int_{q_{min}}^{q_{max}} \left[ 16\pi^4 \left(\frac{h_0}{L_0}\right)^2 q^4 + \eta f''(\bar{z}) \left(\frac{L_0}{h_0}\right)^2 \right]^{-1} q\,dq$$

$$\approx f'(\bar{z}) + \frac{f'''(\bar{z})}{8\pi\sqrt{\eta f''(\bar{z})}} \frac{k_B T}{D} \arctan\left( \frac{4\pi^2}{\sqrt{\eta f''(\bar{z})}} \left(\frac{h_0}{b}\right)^2 \right) \quad , \quad (22)$$

where we have taken $q_{max} = L_0/b$ and $q_{min} = 1$. Note that the traction is independent of the domain size $L_0$ as long as $L_0 \gg h_0$, but weakly depends on the choice of the cut-off length $b$. The cut-off length is often taken as a few times of the bond length ($r_0 \sim 0.14$ nm), which is close to the typical values for $h_0$ (~0.3 nm). For convenience, we take $b = h_0$ in subsequent calculations. It is found that the results from Eq. (22) are in close agreement with the summation in Eq. (21) for $L_0/h_0 > 10$.

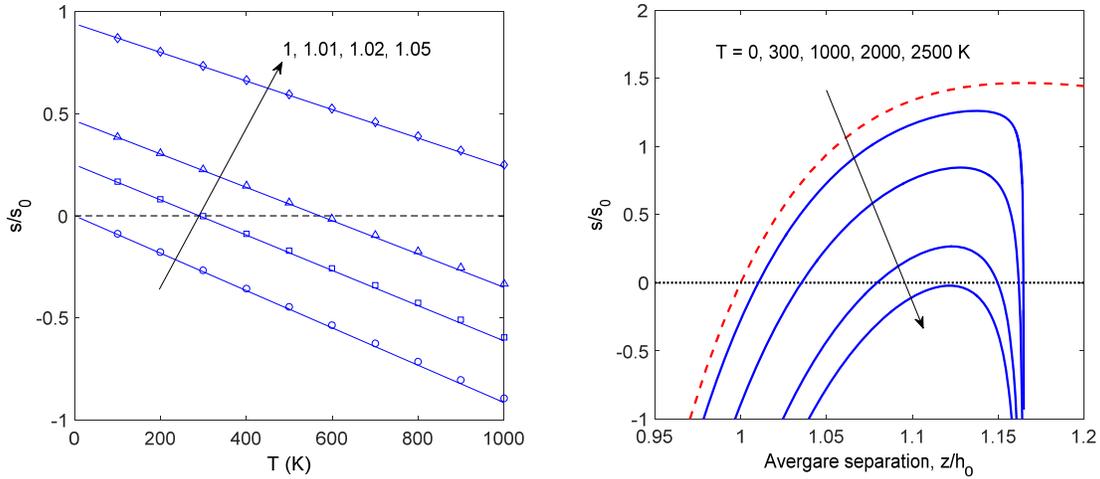

Figure 2. (a) Predicted normal traction as a function of temperature at different average separations, $\bar{z} = 1$, 1.01, 1.02, and 1.05 (symbols by summation and lines by integral approximation); (b) Predicted traction-separation relations at different temperatures, in comparison with the relation at T = 0 K (dashed line). Parameters: D = 1.4 eV, $\Gamma_0$ = 0.242 J/m$^2$, $h_0$ = 0.316 nm, $\eta$ = 0.11, and $\varepsilon_0$ = 0.

As shown in Fig. 2a, the normalized traction decreases linearly with increasing temperature; the linear dependence is expected as a result of the harmonic approximation in the present analysis. For $\bar{z} > 1$, the traction is positive (attraction) at low temperatures but may become negative (repulsion) at high temperatures. Evidently, the entropic effect of thermal rippling leads to an effective repulsion in addition to the van der Waals forces. Figure 2b shows the predicted traction-separation relations at different temperatures. As the temperature increases, the maximum traction (a.k.a., interfacial strength) decreases. In other words, the attractive forces between graphene and substrate are weakened by the entropic repulsion due to thermal rippling. Above a critical temperature ($T_c \sim 2462$ K for $\eta = 0.11$), the traction becomes all repulsive ($s < 0$), meaning that the van der Waals forces are no longer sufficient to keep the graphene attached to the substrate. Moreover, the predicted traction-separation relation is limited by the condition, $f''(\bar{z}) > 0$ or equivalently $\bar{z} < 1.165$. For $\bar{z} > 1.165$, the integral in Eq. (22) is unbounded and the harmonic analysis yields no meaningful result.

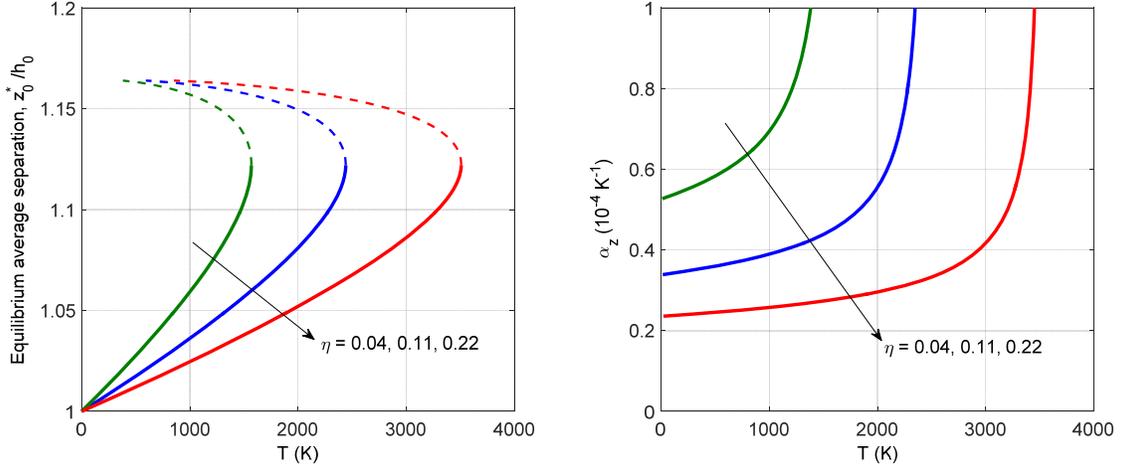

Figure 3. (a) Predicted equilibrium average separation as a function of temperature, with an unstable branch for the critical separation (dashed lines); (b) Predicted out-of-plane coefficient of thermal expansion as a function of temperature.

By setting the traction in Eq. (22) to zero we obtain two equilibrium average separations (see Fig. 2b), one is stable with $\left(\frac{\partial s}{\partial \bar{z}}\right)_T > 0$ at $\bar{z} = \bar{z}_0^*(T)$ and the other is unstable with $\left(\frac{\partial s}{\partial \bar{z}}\right)_T < 0$ at $\bar{z} = \bar{z}_c(T)$. The latter is called the critical average separation, beyond which the traction becomes repulsive by the harmonic analysis. As shown in Fig. 3a, the stable equilibrium average separation increases with temperature almost linearly up to 1000 K, beyond which it becomes nonlinear, and no solution can be found above the critical temperature ($T_c \sim 2462$ K for $\eta = 0.11$). Meanwhile, the critical average separation decreases with increasing temperature (dashed lines in Fig. 3a). At the critical temperature, the two average separations converge at $\bar{z}_0^*(T_c) = 1.122$; hence, by Eq. (22), the critical temperature depends on the van der Waals interactions approximately as $k_B T_c \sim D\sqrt{\eta}$.

The increase of the stable equilibrium average separation with temperature is similar to the out-of-plane thermal expansion of graphite, with a temperature-dependent, positive coefficient of thermal expansion [38]. Quantitatively, the coefficient of out-of-plane thermal expansion (CTE) for the graphene/substrate interface may be defined as $\alpha_z = d\bar{z}_0^*/dT$, which depends on the van der Waals interactions through the dimensionless group $\eta$. As shown in Fig. 3b, the CTE decreases as $\eta$ increases. For $\eta = 0.11$ and $T < 1000$ K, we obtain $\alpha_z \approx 3.5 \times 10^{-5}$ K$^{-1}$, which is slightly larger than the measured out-of-plane CTE of graphite at around 1000 K [39]. The predicted CTE increases with increasing temperature, in qualitative agreement with the measured CTE for graphite. However, the present prediction appears to overestimate the CTE at low temperatures ($T < 200$ K) and at very high temperatures ($T > 2000$ K).

At the equilibrium average separation $\bar{z}_0^*$ for $\varepsilon_0 = 0$, the average rippling amplitude can be obtained from Eq. (12) in form of a discrete summation as

$$\frac{\langle \delta^2 \rangle}{h_0^2} = \frac{k_B T}{D} \sum_{i,j=-n}^{n} \left[ 16\pi^4 (i^2 + j^2)^2 \left(\frac{h_0}{L_0}\right)^2 + \eta f''(\bar{z}_0^*)\left(\frac{L_0}{h_0}\right)^2 \right]^{-1}, \qquad (23)$$

For $n \to \infty$, the summation can be evaluated by an integral approximation and the root-mean-square (RMS) amplitude of thermal rippling is then obtained as

$$\bar{\delta} = \sqrt{\langle \delta^2 \rangle} \approx h_0 \sqrt{\frac{k_B T}{8D}} \left[\eta f''(\bar{z}_0^*)\right]^{-1/4}. \qquad (24)$$

For a freestanding membrane, the rippling amplitude can be obtained from Eq. (23) with $\eta = 0$, which recovers the result in the previous study [10]:

$$\bar{\delta} \approx \frac{L_0}{4\pi^{3/2}} \sqrt{\frac{k_B T}{D}}. \qquad (25)$$

Apparently, as a result of the harmonic approximation, the rippling amplitude of a freestanding membrane scales linearly with the domain size ($L_0$), although a power-law scaling was observed in MD simulations due to anharmonic effects [10]. In contrast, with the presence of van der Waals interactions ($\eta > 0$), the rippling amplitude in Eq. (24) is independent of the domain size (for $L_0 \gg h_0$). Figure 4a shows the predicted rippling amplitude as a function of temperature for different values of $\eta$. Evidently, comparing to the freestanding graphene, the presence of adhesive interactions considerably suppresses the amplitude of thermal rippling, and the normalized RMS amplitude decreases with increasing $\eta$.

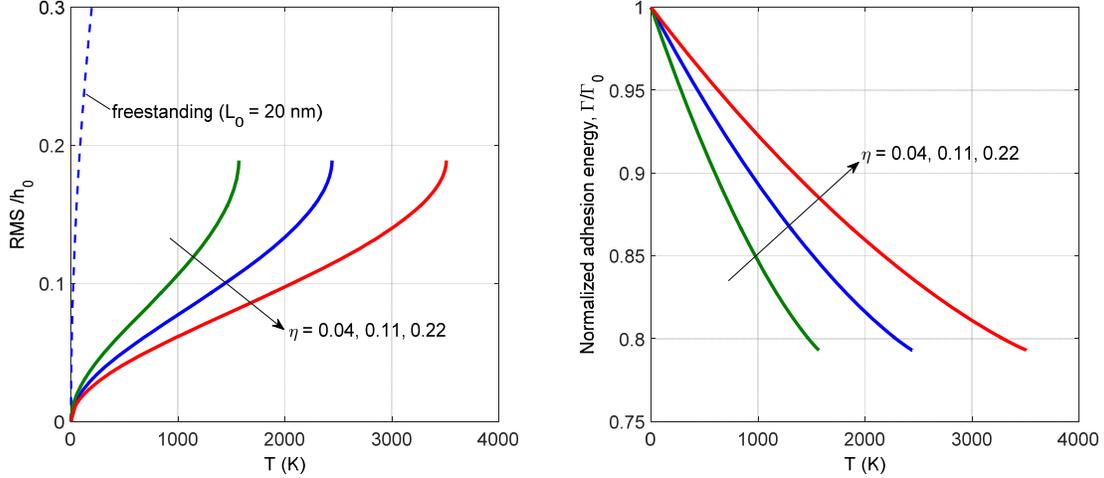

Figure 4. (a) Predicted RMS amplitude of thermal rippling as a function of temperature. (b) Normalized adhesion energy as a function of temperature due to the effect of thermal rippling.

The effective adhesion energy may be defined as the difference between the Helmholtz free energy at the equilibrium average separation ($\bar{z} = \bar{z}^*$) and that at infinite separation ($\bar{z} \to \infty$). For $\varepsilon_0 = 0$, we have

$$\Gamma(T) = -\frac{1}{L_0^2}\left[A(\bar{z}_0^*,0,T) - A(\infty,0,T)\right]$$
$$\approx -V(\bar{z}_0^* h_0) - \frac{k_B T}{L_0^2} \sum_{k(\mathbf{q}_k \cdot \mathbf{e}_y \geq 0)} \ln\left(1 + \frac{V''(\bar{z}_0^* h_0)}{D q_k^4}\right).$$
(26)

It can be seen that, as $T \to 0\,\text{K}$, we have $\bar{z}_0^* \to 1$ and $\Gamma \to \Gamma_0 = -V(h_0)$ as expected from the interaction energy function in Eq. (1). By integral approximation we obtain

$$\frac{\Gamma(T)}{\Gamma_0} \approx -f(\bar{z}_0^*) - \frac{k_B T}{8D}\sqrt{\frac{f''(\bar{z}_0^*)}{\eta}}.$$
(27)

As shown in Fig. 4b, the adhesion energy decreases with increasing temperature, almost linearly up to about 1000 K. Interestingly, while the statistical effect of thermal rippling leads to an effective repulsion and hence an effectively lower adhesion energy with increasing temperature, an opposite effect was predicted by considering the electromagnetic modes and thermal radiation, where the attractive van der Waals forces increase with increasing temperature [29-32]. For the case of an atomic monolayer interacting with a solid substrate, the two effects may co-exist, leading to a more complicated dependence on temperature. Without considering the increasing attractive van der Waals forces, the entropic effect is overestimated by the thermal rippling effect alone. On the other hand, the out-of-plane CTE of graphite was underestimated by the first-principle calculations with a quasiharmonic approximaiton [38], possibly because the thermal rippling effects were not fully taken into account. Thus, the coupling of the two competing effects would be of interest for further studies.

Alternatively, the predicted traction-separation relations (see Fig. 2b) may be used to determine the adhesion energy (or work of separation), by integrating the traction from the equilibrium average separation ($\bar{z}^*$) to the critical average separation ($\bar{z}_c$). This is equivalent to the difference in the Helmholtz free energy at the two equilibrium separations, which would give a much lower adhesion energy due to the much shorter range of separation ($\bar{z} < \bar{z}_c < 1.165$) accessible by the harmonic analysis.

The effect of pre-strain on the interfacial traction-separation relation is shown in Fig. 5a, where the summation in Eq. (21) is calculated by an integral approximation similar to that in Eq. (22). When $\varepsilon_0 > 0$, we obtain

$$\frac{s}{s_0} \approx f'(\bar{z}) + \frac{k_B T}{4\pi D} \frac{f'''(\bar{z})}{\sqrt{4\eta f''(\bar{z}) - \beta^2 \varepsilon_0^2}} \left[ \frac{\pi}{2} - \arctan\left(\frac{\beta \varepsilon_0}{\sqrt{4\eta f''(\bar{z}) - \beta^2 \varepsilon_0^2}}\right) \right] \quad (28)$$

if $4\eta f''(\bar{z}) > \beta^2 \varepsilon_0^2$, or

$$\frac{s}{s_0} \approx f'(\bar{z}) + \frac{k_B T}{8\pi D} \frac{f'''(\bar{z})}{\sqrt{\beta^2 \varepsilon_0^2 - 4\eta f''(\bar{z})}} \ln\left(\frac{\beta \varepsilon_0 + \sqrt{\beta^2 \varepsilon_0^2 - 4\eta f''(\bar{z})}}{\beta \varepsilon_0 - \sqrt{\beta^2 \varepsilon_0^2 - 4\eta f''(\bar{z})}}\right) \quad (29)$$

if $4\eta f''(\bar{z}) < \beta^2 \varepsilon_0^2$. When $\varepsilon_0 < 0$, the traction is unbounded if $4\eta f''(\bar{z}) < \beta^2 \varepsilon_0^2$ and only the result for $4\eta f''(\bar{z}) > \beta^2 \varepsilon_0^2$ is meaningful. Notably, the traction-separation relation depends on the pre-strain sensitively when $\varepsilon_0 < 0$, with decreasing strength for increasingly large compressive strain. This again can be attributed to the effect of entropic repulsion due to thermal rippling that is amplified by the compressive strain. Beyond a critical compressive strain, the traction becomes all repulsive. On the other hand, when $\varepsilon_0 > 0$, the entropic repulsion is reduced so that the maximum traction increases with increasing strain, slowly approaching the limit at $T = 0$ K (dashed line).

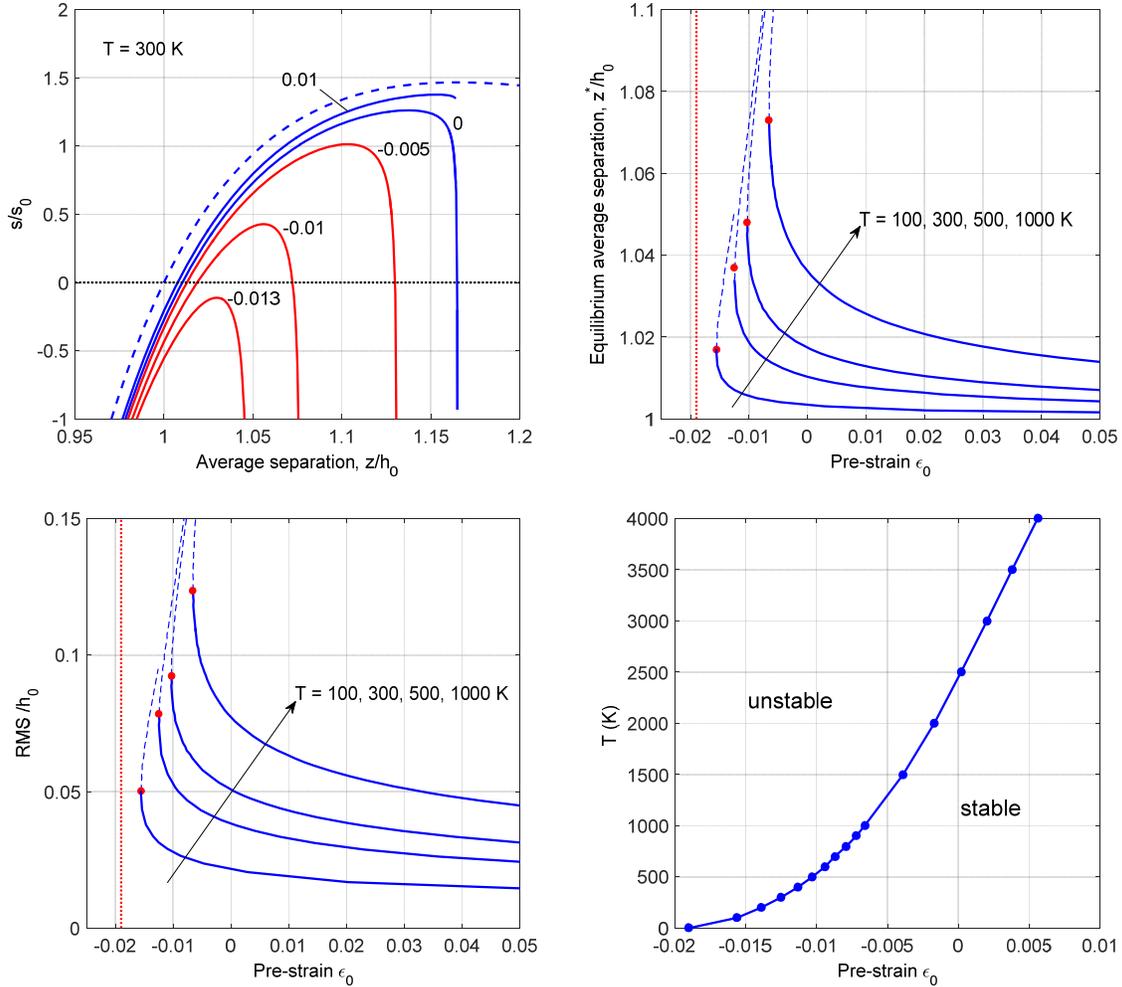

Figure 5. Effects of pre-strain by the statistical mechanics analysis (with parameters: E* = 403 N/m, D = 1.4 eV, $\Gamma_0$ = 0.242 J/m$^2$, h$_0$ = 0.316 nm, η = 0.11). (a) Traction-separation relations at T = 300 K with different pre-strains as indicated. The dashed line is the traction-separation relation at T = 0 K, independent of the pre-strain. (b) Equilibrium average separation, with a critical strain at each temperature. (c) RMS amplitude of thermal rippling. (d) Critical strain versus temperature (a stability phase diagram).

By setting the interfacial traction to zero, we obtain the equilibrium average separation $\bar{z}^*(\varepsilon_0, T)$ as a function of the pre-strain at different temperatures, as shown in Fig. 5b. Similar to Fig. 3a, there are two branches for the equilibrium separation at each temperature, one stable and the other unstable (critical average separation, $\bar{z}_c(\varepsilon_0, T)$, shown as dashed lines). The two branches converge at a critical strain ($\varepsilon_c$), below which no solution can be found as the traction becomes all repulsive. Correspondingly, Fig. 5c shows the effect of pre-strain on the rippling amplitude. By Eq. (12) and Eq. (15), the rippling amplitude at the equilibrium average separation is obtained as

$$\bar{\delta} = h_0 \sqrt{-\frac{2f'(\bar{z}^*)}{f'''(\bar{z}^*)}}. \tag{30}$$

The rippling amplitude decreases with a tensile pre-strain and increases with a compressive strain. As a tensile strain tends to reduce the amplitude of thermal rippling, it reduces the entropic repulsion and hence the equilibrium average separation (Fig. 5b). The opposite is true for a compressive strain until it reaches the critical strain ($\varepsilon_c$). The rippling amplitude increases rapidly near the critical strain, resembling a buckling instability. Beyond the critical strain ($\varepsilon_0 < \varepsilon_c$), a nonlinear analysis with anharmonic effects would be necessary for further studies. The critical strain as predicted by the present analysis depends on temperature through the dimensionless group, $k_B T/D$. In addition, it depends on the van der Waals interactions and the mechanical properties of graphene through two other dimensionless groups, $\eta = \Gamma_0 h_0^2/D$ and $\beta = E^* h_0^2/D$. As $T \to 0\,\mathrm{K}$, the critical strain approaches the buckling strain, $\varepsilon_B = -6\sqrt{3\eta}/\beta$; the latter was predicted previously by Aitken and Huang [27] without considering the effect of thermal rippling. At a finite temperature, with thermal rippling, the critical strain becomes less compressive, i.e., $\varepsilon_c > \varepsilon_B$, as shown in Fig. 5d. At very high temperatures, the membrane could be unstable even under a tensile strain (e.g., $\varepsilon_c > 0$). The critical temperature noted in Fig. 3a is simply the temperature with a zero critical strain ($\varepsilon_c = 0$). Hence, Fig. 5d may be considered as a stability phase diagram in terms of temperature and pre-strain.

By Eq. (20), the average in-plane stress in graphene is obtained with an entropic contribution (the rippling stress) as

$$\begin{aligned}\tilde{\sigma}^*(\varepsilon_0, T) &= E^* \frac{k_B T}{16\pi^2 D} \sum_{i,j} \left[(i^2+j^2) + \frac{E^* \varepsilon_0 L_0^2}{4\pi^2 D} + \frac{1}{i^2+j^2} \frac{\eta f''(\bar{z}^*)}{16\pi^4} \frac{L_0^4}{h_0^4}\right]^{-1} \\ &\approx E^* \left[\frac{k_B T}{32\pi D} \ln\left(1 + \frac{4\pi^2}{\eta f''(\bar{z}^*)} \frac{E^* \varepsilon_0 h_0^2}{D}\left(\frac{h_0}{b}\right)^2 + \frac{16\pi^4}{\eta f''(\bar{z}^*)}\left(\frac{h_0}{b}\right)^4\right) + \frac{E^* \varepsilon_0 h_0^2}{4D} \frac{f'(\bar{z}^*)}{f'''(\bar{z}^*)}\right].\end{aligned} \tag{31}$$

Figure 6a shows that the entropic rippling stress increases with increasing temperature, but decreases with increasing pre-strain, following the same trend as the rippling amplitude (Fig. 5c).

The total stress, with the effect of in-plane thermal expansion, is shown in Fig. 6b as a function of pre-strain for $T = 1000$ K. Here we have assumed that the biaxial modulus $E^*$ of graphene is independent of temperature and strain. Due to in-plane thermal expansion, the in-plane stress-strain relation simply shifts downward at a finite temperature before the rippling stress is taken into account. With thermal rippling, the total stress becomes more tensile with a slightly nonlinear dependence on the pre-strain. The effective modulus, defined as the slope of the stress-strain curve, is lower than $E^*$ and depends on temperature, similar to the effective modulus for a freestanding graphene as discussed in the previous study [10].

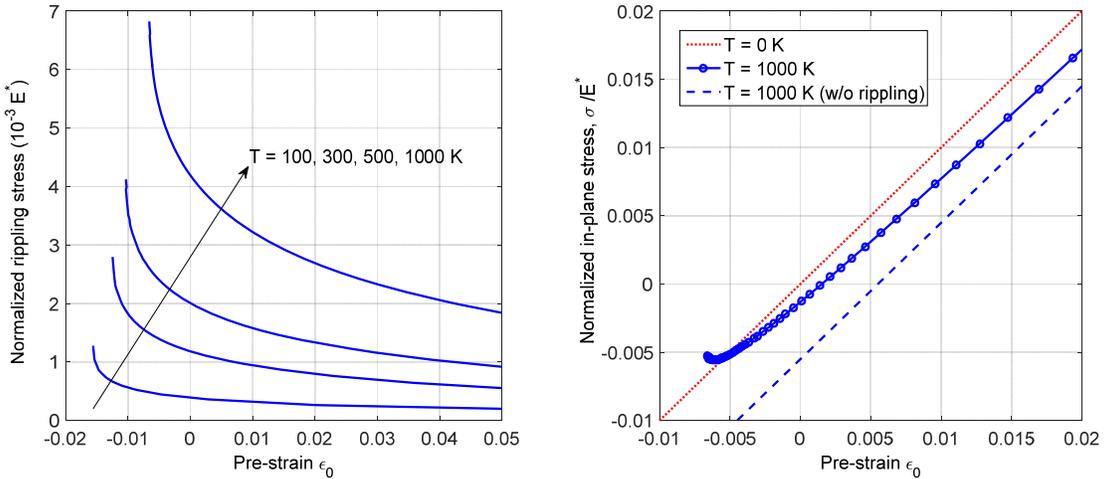

Figure 6. (a) Predicted rippling stress as a function of pre-strain; (b) Comparison of the average in-plane stresses at 1000 K with and without rippling.

## III. MOLECULAR DYNAMICS SIMULATION

The theoretical predictions by the continuum statistical mechanics analysis in Section II are compared to MD simulations using LAMMPS [40]. A square-shaped graphene membrane ($L_0 \sim$ 20 nm) is placed on top of a flat surface as a rigid substrate. The van der Waals interactions between the carbon atoms of graphene and the substrate are specified in form of Eq. (1) with two parameters ($\Gamma_0$ and $h_0$). The energy per unit area is converted to energy per atom by using the area per atom in the graphene lattice, $A_0 = \frac{3}{4}\sqrt{3} r_0^2$, where $r_0 = 0.142$ nm for the bond length. The dimensionless parameter $\eta$ is varied by changing the reference adhesion energy $\Gamma_0$ and equilibrium separation $h_0$. Here, we use two different values for $h_0$: 0.316 and 1.0 nm; the former is predicted by DFT calculations for graphene on $SiO_2$ [23], while the latter is taken as an upper bound from measurements [17, 41, 42]. The value of $\Gamma_0$ is varied between 0.1 and 1.0 J/m$^2$, as the typical range for the adhesion energy from both experiments and theoretical calculations [17-23].

The second-generation reactive empirical bond-order (REBO) potential [43] is used for the carbon-carbon interactions in graphene. With the REBO potential, the mechanical properties of graphene in the ground state ($T = 0$ K) have been predicted previously [44-46]: $E = 243$ N/m, $\nu = 0.397$, and $D = 1.4$ eV. Although these values are different from DFT calculations [47, 48], they are used in the present study to compare the theoretical predictions with the MD simulations. Under an equi-biaxial pre-strain ($\varepsilon_0$), the theoretical results depend on a dimensionless group, $\beta = E^* h_0^2 / D$. Despite the discrepancy in the 2D Young's modulus and Poisson's ratio, the biaxial modulus, $E^* = E/(1-\nu)$, predicted by the REBO potential ($E^* \sim 403$ N/m) is in close agreement with DFT ($E^* \sim 406$ N/m). The bending modulus ($D$) is also in close agreement with DFT (~1.5 eV) [45, 47].

MD simulations are performed in NVT ensemble with periodic boundary conditions, where the temperature is controlled by the Nose-Hoover thermostat. The equi-biaxial pre-strain $\varepsilon_0$ is applied to the graphene membrane by simultaneously changing the two in-plane dimensions as $L = L_0(1+\varepsilon_0)$, where $L_0$ is the side length of the square-shaped membrane in the ground state ($T = 0$ K). It is found that the simulation results are independent of the membrane size as long as $L_0 \gg h_0$, and only the simulations with $L_0 \approx 20$ nm (see Fig. 1) are presented. Periodic boundary conditions are applied in all three directions. The thickness dimension of the simulation box is set to be 10 nm so that it is large enough to avoid interactions between periodic images. Each simulation runs up to 40 ns with a time step of 1 fs. The first 10 ns is for the system to equilibrate with the prescribed temperature and pre-strain, and the subsequent 30 ns is used for calculating the time-averaged quantities.

The normalized equilibrium average separation is calculated for each MD simulation as

$$\bar{z}^*(T, \varepsilon_0, \eta) = \frac{1}{Nh_0} \left\langle \sum_{i=1}^{N} z_i \right\rangle_t, \qquad (32)$$

where $N$ is the total number of carbon atoms and $\langle z_i \rangle_t$ is the time-averaged $z$-coordinate of the $i$-th atom. The mean amplitude of the out-of-plane thermal rippling is calculated by a time-averaged RMS, namely

$$\bar{\delta}(T,\varepsilon_0,\eta) = \sqrt{\frac{1}{N}\left\langle \sum_{i=1}^{N}(z_i - \bar{z}^* h_0)^2 \right\rangle_t}, \qquad (33)$$

The average in-plane stress in graphene is evaluated by the time-averaged 2D virial stress:

$$\boldsymbol{\sigma} = \frac{1}{L^2}\left\langle \frac{1}{2}\sum_{\substack{i,j \\ i\neq j}} \mathbf{F}_{ij} \otimes (\mathbf{r}_j - \mathbf{r}_i) - \sum_i m_i \mathbf{v}_i \otimes \mathbf{v}_i \right\rangle_t, \qquad (34)$$

where $\mathbf{F}_{ij}$ is the interatomic force between two carbon atoms ($i$ and $j$), $\mathbf{r}_i$ is the position vector of $i$-th atom, $\mathbf{v}_i$ is the velocity vector, and $m_i$ is the atomic mass.

Finally, the time-averaged interaction potential energy (per unit area) is calculated, for which the corresponding ensemble average can be predicted by the statistical mechanics analysis as

$$\frac{\langle U_I \rangle}{L_0^2} \approx V(\bar{z}^* h_0) + \frac{1}{2}V''(\bar{z}^* h_0)\langle \delta^2 \rangle = \Gamma_0\left(f(\bar{z}^*) - \frac{f'(\bar{z}^*)f''(\bar{z}^*)}{f'''(\bar{z}^*)}\right). \qquad (35)$$

We note that the average interaction energy differs from the effective adhesion energy defined by the Helmholtz free energy (Eq. 26). The latter may be calculated by the steered MD simulations [49], which is left for future studies.

## IV. RESULTS AND DISCUSSION

In this section we compare the theoretical predictions by the statistical mechanics analysis in Section II with the MD simulations. First, we compare the RMS amplitude of thermal rippling (Fig. 7a) and the equilibrium average separation (Fig. 7b) for cases with zero pre-strain ($\varepsilon_0 = 0$). Both increase with increasing temperature as a result of the entropic effect. Increasing the adhesion energy ($\eta$) reduces the rippling amplitude and hence the entropic repulsion, leading to less expansion in the equilibrium separation. The results from MD simulations agree reasonably well with the theoretical predictions at relatively low temperatures. At high temperatures the statistical mechanics analysis over-predicts the amplitude of thermal rippling, possibly due to the harmonic approximation. The predicted critical temperature is not observed in the MD simulations. It is possible that the anharmonic effects not considered in the present analysis are substantial at high temperatures, suppressing the rippling amplitude and delaying the critical temperature behavior. Figure 7c shows the average interaction energy between graphene and substrate, decreasing with increasing temperature. The same trend is predicted for the effective adhesion energy (Fig. 4b). By Eq. (35), the normalized interaction energy with $U_0 = -\Gamma_0 L_0^2$ depends on the rippling amplitude and the average separation. Again, the theoretical prediction agrees with the MD simulations at relatively low temperatures.

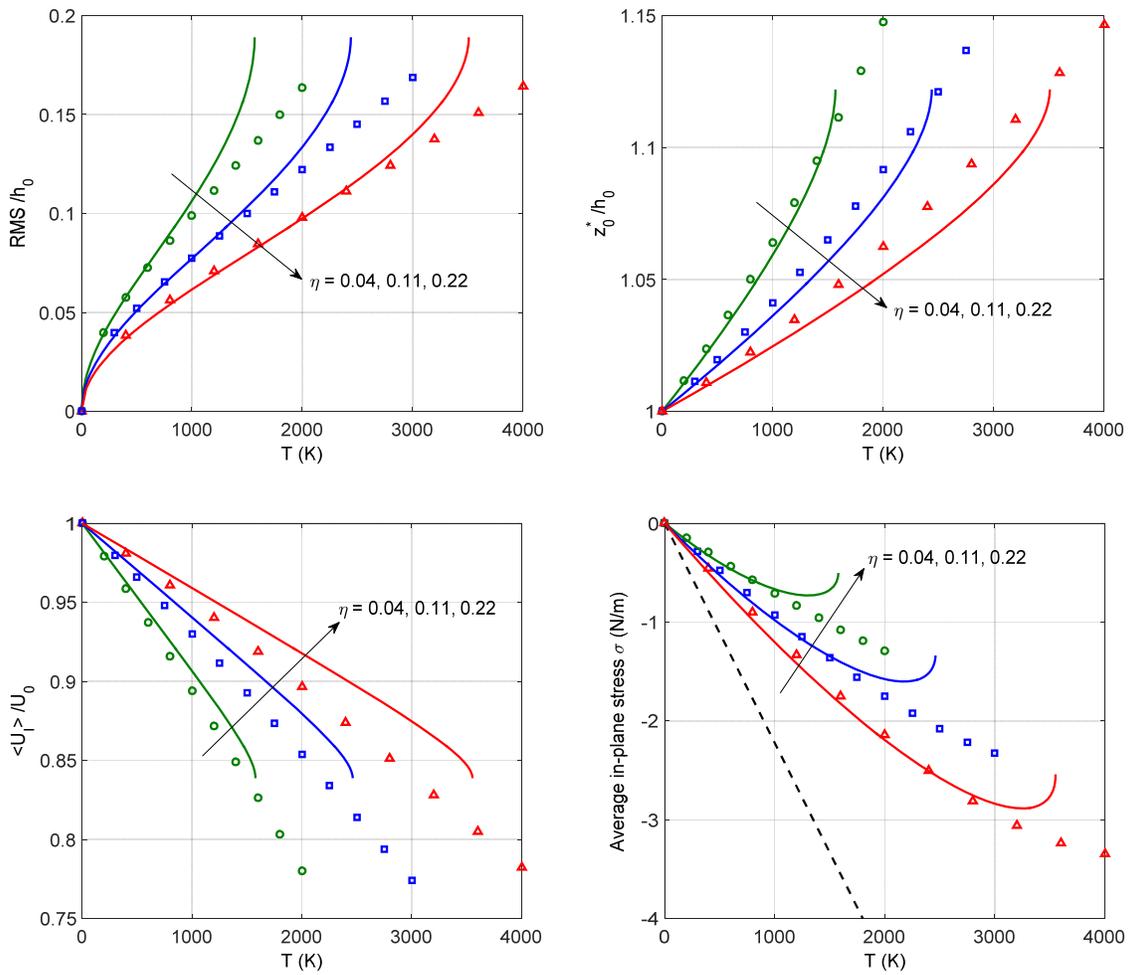

Figure 7. Comparison between theoretical predictions and MD ($\varepsilon_0 = 0$): (a) RMS amplitude of thermal rippling as a function of temperature for different $\eta$. (b) Equilibrium average separation as a function of temperature. (c) Average interaction energy between graphene and substrate. (d) Average in-plane stress in graphene (dashed line for the case of no rippling). All symbols are from MD simulations and lines by the theoretical predictions.

Constrained at zero pre-strain (relatively to the ground state at 0 K), a thermal stress is induced in graphene at a finite temperature. By Eq. (20), the amplitude of thermal stress would increase linearly with temperature if the out-of-plane rippling is completely suppressed, as shown by the dashed line in Fig. 7d. While the positive 2D-CTE ($\alpha_{2D} > 0$) leads to a compressive thermal stress, the rippling stress $\tilde{\sigma}^*$ is tensile, as predicted by Eq. (31) and shown in Fig. 6a. As a result, the average thermal stress in graphene becomes less compressive and depends on the adhesive interactions with the substrate. In contrast, for a freestanding graphene the thermal stress was found to be tensile due to significantly larger rippling stress [10]. Figure 7d shows that the thermal stresses obtained from MD simulations agree reasonably well with the theoretical prediction at relatively low temperatures.

The effects of pre-strain are compared in Fig. 8. First, the rippling amplitudes at four different temperatures are shown with pre-strains ranging from -0.02 to 0.06 (Fig. 8a). The results from MD simulations agree well with the predictions for the cases with a tensile pre-strain ($\varepsilon_0 > 0$). The statistical mechanics analysis predicts a temperature dependent critical strain (Fig. 5d), beyond which the harmonic approximation yields no meaningful result. The RMS amplitude of thermal rippling from MD simulations increases dramatically as the pre-strain changes from -0.01 to -0.02, indicating a critical strain in between. Figure 9 shows the morphology of the supported graphene at 300 K with a pre-strain of -0.02, where a zigzag buckling pattern is observed. Similar buckling patterns are observed at other temperatures. Such a buckling phenomenon resembles the telephone cord blistering in thin films as a result of biaxial compression and interfacial delamination [50]. Apparently, the largest separation shown in Fig. 9 is greater than 1 nm (~3$h_0$), for which the van der Waals interactions with the substrate become negligible and the graphene may be considered as delaminated locally from the substrate. A few recent studies have also simulated buckling of substrate-supported graphene with a variety of morphological patterns such as wrinkles, folds, and crumpling [51-53]. However, the transition from thermal rippling to buckling is noted for the first time in the present study. We leave it for further studies to determine the critical strain for this transition and its dependence on the interfacial adhesion and temperature.

The equilibrium average separation as a function of the pre-strain is compared in Fig. 8b. The trend is similar to the rippling amplitude because the entropic repulsion increases with increasing rippling amplitude. The average separation becomes much larger (~1.4 $h_0$) at $\varepsilon_0 = -0.02$, a result of the rippling to buckling transition. Similarly, the average interaction energy is compared in Fig.

8c, which decreases as both the rippling amplitude and the equilibrium average separation increase under a compressive strain. After the buckling transition, the average interaction energy drops dramatically to less than 80% of the reference value ($U_0 = -\Gamma_0 L_0^2$) at $\varepsilon_0 = -0.02$. The comparisons in Fig. 8 (a-c) show that the theoretical predictions in Section II are reasonable for the cases with relatively low temperatures (T < 1000 K) and subcritical pre-strains ($\varepsilon_0 > -0.01$).

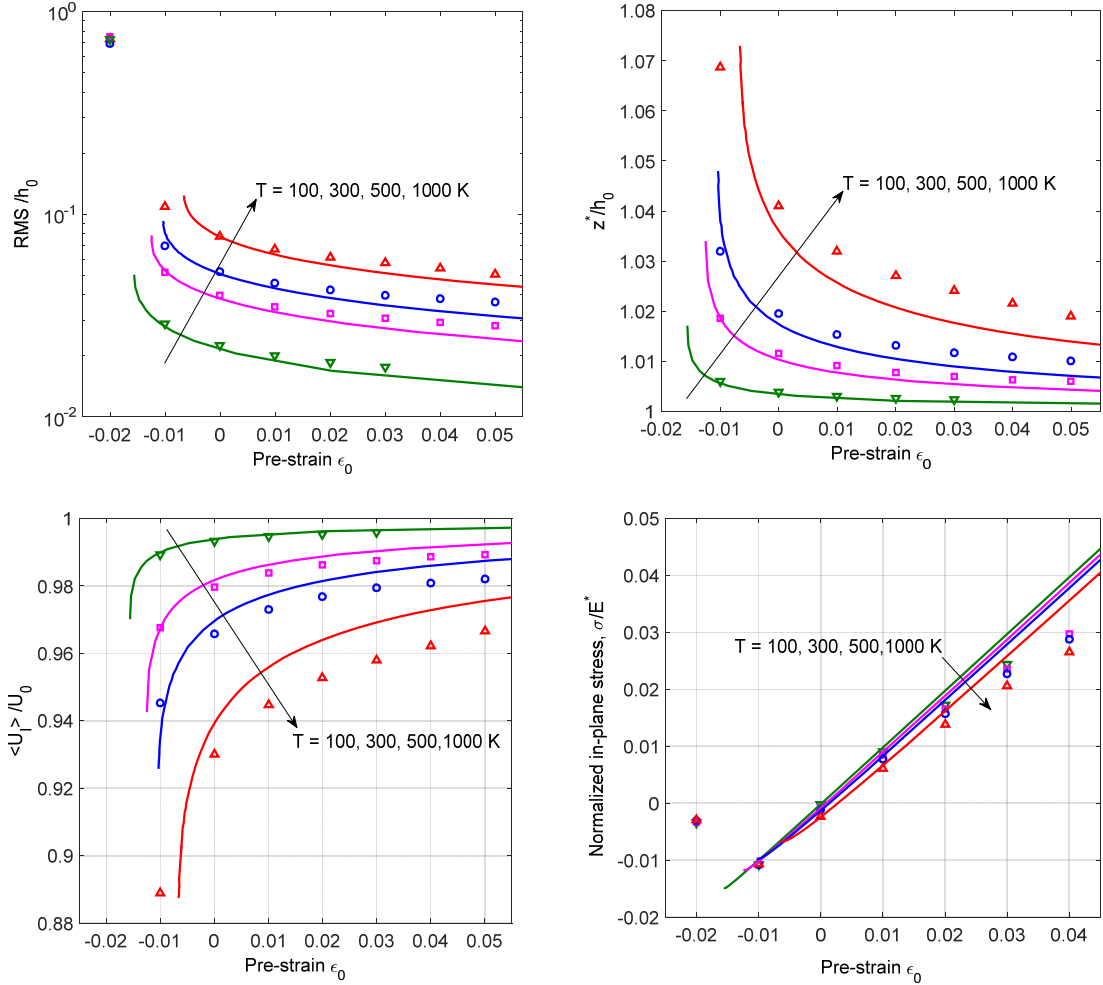

Figure 8. Effects of pre-strain by MD ($\Gamma_0$ = 0.242 J/m$^2$, $h_0$ = 0.316 nm, and η = 0.11). (a) RMS amplitude of thermal rippling as a function of strain at different temperatures. (b) Equilibrium average separation as a function of strain. (c) Average interaction energy between graphene and substrate. (d) Average in-plane stress in graphene. All symbols are from MD simulations and lines by the theoretical predictions.

Figure 8d compares the normalized in-plane stress of graphene. As noted in Fig. 6d, the average in-plane stress is subject to two competing effects. Relative to the stress-strain relation at 0 K, the stress becomes more compressive at a finite temperature (T > 0 K) due to the positive lattice expansion but becomes less compressive due to thermal rippling. The two effects combine to give a weak temperature dependence for the in-plane stress-strain relation of the supported graphene. The results from MD simulations agree with the theoretical predictions when the strain is small ($-0.01 < \varepsilon_0 < 0.01$) for temperatures up to 1000 K. At larger tensile strains ($\varepsilon_0 > 0.01$), the stresses from MD simulations are lower because of the intrinsic elastic nonlinearity of graphene as discussed in previous studies [10, 44, 48]. At larger compressive strains (e.g., $\varepsilon_0 = -0.02$), the compressive stress is largely relaxed and nearly independent of temperature after the buckling transition.

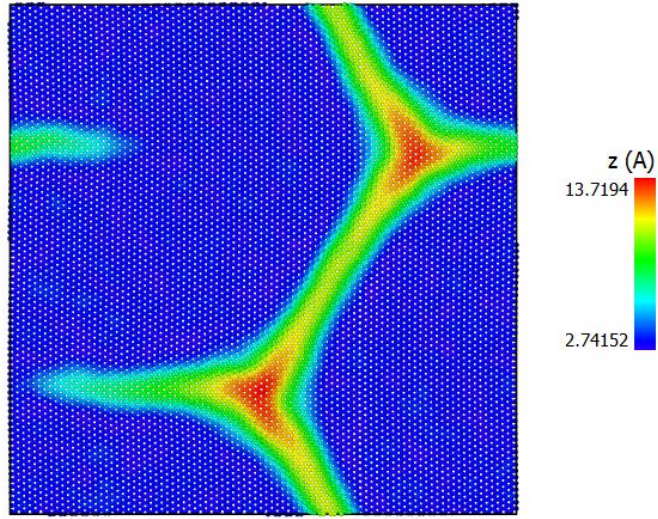

Figure 9. Buckling of a substrate-supported graphene by MD simulation at 300 K with a biaxial pre-strain of -0.02. The side length of the graphene membrane as shown is about 20 nm, and the interfacial properties are: $\Gamma_0$ = 0.242 J/m$^2$ and $h_0$ = 0.316 nm.

We close this section by commenting on the major differences between substrate-supported graphene and freestanding graphene. For freestanding graphene ($\eta = 0$), as shown in the previous study [10], the rippling amplitudes from MD simulations are considerably lower than the predictions by the harmonic analysis (even at low temperatures) and depend on the size of the graphene membrane by a power law instead of the linear scaling predicted by the harmonic approximation. For supported graphene with adhesive interactions ($\eta > 0$), the rippling amplitudes are independent of the membrane size, and the harmonic approximation becomes more applicable since the rippling amplitude is much smaller than freestanding graphene. The comparisons in Figs. 7 and 8 suggest that the theoretical predictions by the harmonic approximations are reasonable as long as the rippling amplitude is relatively small (e.g., $\bar{\delta}/h_0 < 0.1$). For the case with $\eta = 0.11$, the applicable temperature range is up to 1000 K with the pre-strain $\varepsilon_0 > -0.01$.

## V. SUMMARY

Thermal rippling of a substrate-supported graphene depends on the adhesive interactions between graphene and the substrate, and the statistical nature of thermal rippling leads to an entropic effect on the graphene-substrate interactions. This inter-relationship between thermal rippling and adhesion is theoretically analyzed by a continuum statistical mechanics approach under harmonic approximations. Comparisons with MD simulations show that the theoretical predictions on the rippling amplitude, the equilibrium average separation, and the average interaction energy are in reasonable agreement at relatively low temperatures, when the rippling amplitude is relatively small. Of particular interest is the entropic effects of thermal rippling that lead to an effective repulsion, and as a result, the equilibrium average separation increases and the effective adhesion energy decreases with increasing temperature. Moreover, the presence of a biaxial pre-strain in graphene could either reduce or amplify the thermal rippling and the entropic effects, depending on the sign of strain (tensile or compressive). A rippling-to-buckling transition is predicted and observed in MD simulations beyond a critical compressive pre-strain. These theoretical and numerical results shed light on the commonly observed morphological features (wrinkles, buckles, and folds) in substrate-supported graphene and other 2D materials, and in particular, on the effects of adhesive interactions and temperature. Further studies would extend the statistical mechanics analysis to account for the anharmonic effects and consider more realistic substrate surfaces with roughness.